\documentclass[journal=jpclcd,manuscript=letter,layout=twocolumn]{achemso}

\usepackage[version=3]{mhchem} 
\usepackage[T1]{fontenc}       
\usepackage{dcolumn}

\author{Chao Zhang}
\affiliation[]{Department of Chemistry-\AA ngstr\"om Laboratory, Uppsala
  University, L\"agerhyddsv\"agen 1, BOX 538, 75121, Uppsala, Sweden}
\email{chao.zhang@kemi.uu.se}
\author{J\"urg Hutter}
\affiliation[]{Institut f\"ur Chemie, Universit\"at Z\"urich,
  Winterthurerstrasse 190, CH-8057 Z\"urich, Switzerland}
\author{Michiel Sprik}
\affiliation[]{Department of Chemistry, University of Cambridge, Lensfield Rd, Cambridge CB2 1EW, United Kingdom}

\title[]{Coupling of Surface Chemistry and Electric Double Layer at TiO$_2$ Electrochemical Interfaces}

\begin{document}

\begin{tocentry}

\includegraphics[width=5cm]{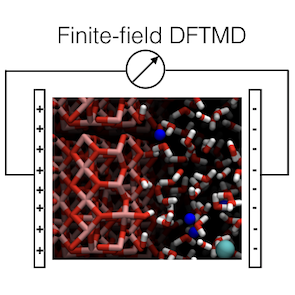}

\end{tocentry}

\begin{abstract}

Surfaces of metal oxides at working conditions are usually electrified due to the  acid-base chemistry. The charged interface compensated with counterions forms the so-called electric double layer. The coupling of surface chemistry and electric double layer is considered to be crucial but poorly understood because of lacking the information at the atomistic scale. Here, we used the latest development in density functional theory based finite-field molecular dynamics simulation to investigate pH-dependence of the Helmholtz capacitance at electrified rutile TiO$_2$ (110)-NaCl electrolyte interfaces. It is found that, due to competing forces from surface adsorption and from  electric double layer, water molecules have a stronger structural fluctuation at high pH and this leads to a much larger capacitance. It is also seen that, interfacial proton transfers at low pH increase significantly the capacitance value. These findings elucidate the microscopic origin for the same trend observed in titration experiments. 
\end{abstract}


The surface chemistry of metal oxides is more involved than that of metals because the surface charge depends on proton exchange driven by pH of the environment~\cite{Ardizzone:1996ca, GRO20191}. The condition in which the surface has no net proton charge is called the point of zero proton charge (PZPC)~\cite{Adekola:2011kk, Zarzycki:2012ds, Kosmulski:2014fv}. However, working conditions for most earth-abundant metal oxide based photo-catalysts are not at PZPC and the corresponding solid-liquid interface is highly electrified~\cite{Lyons:2017cg}. By attracting counterions from the solution side, the electric double layer (EDL) is formed at the interface in which the electric field can be as high as 10$^9$ V/m~\cite{Schmickler:2010th}. Therefore, including the coupling of surface chemistry and EDL at metal oxide-electrolyte interfaces is necessary.

The differential capacitance $C_\text{EDL}$ of the EDL at oxide materials-electrolyte interfaces has been long regarded as a key probe of its structure. For semiconducting oxides, the capacitance can be resolved into three distinct components connected in series~\cite{Nozik:1996gv,Gratzel:2001ub}:

\begin{equation}
\label{C_EDL}
1/C_\text{EDL}=1/C_\text{SC}+1/C_\text{H}+1/C_\text{GC}
\end{equation}

The first component is the result of the depletion or accumulation of electrons in the space charge (SC) layer of the semiconductor electrode, which can be 10 to 100 nm thick. The second component $C_\text{H}$ is the Helmholtz capacitance due to specific adsorption of hydroxide groups or protons and corresponding counter ions. The last component $C_\text{GC}$ called the Gouy-Chapman capacitance, stems from the diffusive electrolyte and depends on the ionic strength.

The relationship between surface structure and ion complexation of oxide-solution interface is a long-standing topic in colloid chemistry. Experimental interest in these metal oxide systems dates back to early 60s~\cite{BOLT:1957dr, Parks:1962eva}. In titration experiments, the variation of surface charges due to adsorption/desorption of H$^+$ is measured against the pH in solution. To interprete experiments, 2-pK model including a phenomenological neutral site (SOH$^0$), dissociate site (SO$^-$) and associate site (SOH$_2^+$) was introduced (``S'' for the surface)~\cite{Parks:1965gaa}. However, the structural information was completely overlooked there until the development of surface complexation model in 1989~\cite{Hiemstra:1989vh, Hiemstra:1989br}. In the multi-site complexation (MUSIC) model and other types of bond valence model~\cite{Bickmore:2004gya, Brown:2009iu},  proton affinity (or pKa) is correlated with the Pauling bond valance charge  (the charge of a cation divided by its coordination number) and the Me-H distance. To connect with titration experiments, a Gouy-Chapman-Stern type EDL model was adopted to match the surface charges $q$ and the potential generated from surface complexation models $\Delta \phi$. Thus, the capacitance $C_\text{H}$ in surface complexation models is a (global) fitting parameter to titration experiments~\cite{BICKMORE2006269, Boily:2014jp}.

A more subtle yet important aspect is that proton affinity itself depends on the capacitance of EDL. Indeed, in the original paper of surface complexation model, authors subtracted Nernstian EDL contribution to pKa of the solution monomers but neglected it for the surface protonation reactions~\cite{Hiemstra:1989vh}.  This was inevitable because the dielectric constant of bulk water is well known as 78 in ambient conditions while the dielectric constant of EDL at oxide-electrolyte interfaces remains a puzzle~\cite{Boamah:2018kn}. 

In parallel to the development by colloid chemists, electrochemists have their own interests in electrified metal oxide-electrolyte interfaces. Since the discovery of the Honda-Fujishima effect in 1972 for TiO$_2$~\cite{FUJISHIMA:1972hc}, metal oxide photo-electrochemistry attracts the perennial interest in water splitting research using solar energy. The focus there was to identify metal oxides having the right band alignment with respect to the water redox potential, for example, the conduction band minimum should be higher than the H$^+$/H$_2$ reduction potential~\cite{Nozik:1996gv,Gratzel:2001ub, Gerischer:1978ig}.  Over the years, experimental characterizations have moved from ultra-high vacuum condition to solid-liquid interface due to the advancement in surface science techniques~\cite{Henderson:2011id, Pang:2013fwa, Hussain:2016bu, Diebold:2017bp,Balajka:2018gw}.

As for colloid chemists, the potential distribution at TiO$_2$ aqueous electrolyte interfaces also interested electrochemists. From the slope of Mott-Schottky plots, the Helmholtz capacitance as a function of pH was determined~\cite{Tomkiewicz:1979ifa}. Despite that the reported median Helmholtz capacitance as 50 $\mu$F/cm$^2$ is surprisingly close to the well-quoted value of 64 $\mu$F/cm$^2$ by colloid chemists~\cite{Ridley:2009ig}, opposite trends in its pH-dependence were seen~\cite{Tomkiewicz:1979ifa, Berube:1968im}. 

Similar to the approximation colloid chemists took in building surface complexation model, the standard Pourbaix diagram in electrochemistry neglects the non-Nernstian contribution of EDL to the proton affinity~\cite{GERISCHER19891005, Gileadi:2011fca}. Therefore, elucidating the microscopic origin for the pH-dependence of Helmholtz capacitance and quantifying its contribution to the surface pKa is a common challenge in both fields. 

Density functional theory based molecular dynamics (DFTMD) is a suitable technique to tackle this challenge which encompasses the electronic, structural and dynamical ingredients on an equal footing. DFTMD modeling of solid-liquid interface has been applied to different areas, such as studying the structure of water (defects) at solid interfaces~\cite{Liu:2010co, Anonymous:2012dr, Liu:2012bd, Bandura:2011fg, Tocci:2014by, vonRudorff:2016cf, Selcuk:2016cb}, spectroscopic modeling of surface-sensitive vibrational signals~\cite{Anonymous:2012fp, Wan:2015id, Anonymous:2016be}, computing the surface acidity~\cite{Cheng:2010gg, Sulpizi:2012ti, Liu:2013he, Churakov:2014bl} and determining the redox potential~\cite{Cheng:2012cj, Cheng:2014jb, Pham:2017bu, Ambrosio:2018gf}.

In spite of progress, electric properties, such as the dielectric constant and the interfacial capacitance which are at the heart of modeling the electrified solid-liquid interface, were usually thought to be beyond the reach of DFTMD. Thanks to the development of constant electric displacement $\bar{D}$ Hamiltonian for the modeling of ferroelectric nanocapacitors by Vanderbilt and co-workers~\cite{Stengel:2009cd}, electric properties became more accessible to DFTMD simulations~\cite{Zhang:2016cl, Zhang:2016ho, Zhang:2016ca, Sayer:2017cw, Zhang:2018cf, Sayer:2019bm}. We are therefore in a position to apply the hybrid constant $\bar{D}$ simulation with PBE functional~\cite{PhysRevLett.77.3865} to the charged rutile TiO$_2$ (110)-NaCl electrolyte interface as implemented in CP2K~\cite{Hutter:2013iea, si}. In Eq.~\ref{C_EDL}, $C_\text{SC}$ is the smallest of three capacitances and will normally dominate in the inverse sum. However, its effect and associated depletion layer can been eliminated at the flatband potential condition by an appropriate bias potential~\cite{GERISCHER19891005, Nozik:1996gv}. At the high ionic strength which is relevant for photoelectrocatalysis, the diffuse ionic layer has a higher capacitance and the inverse $C_\text{GC}$ term can therefore be ignored. Based on these considerations, modeling the Helmholtz capacitance $C_\text{H}$ is what we focus on in this work. 

\begin{figure} [h]
\includegraphics[width=1.0\columnwidth]{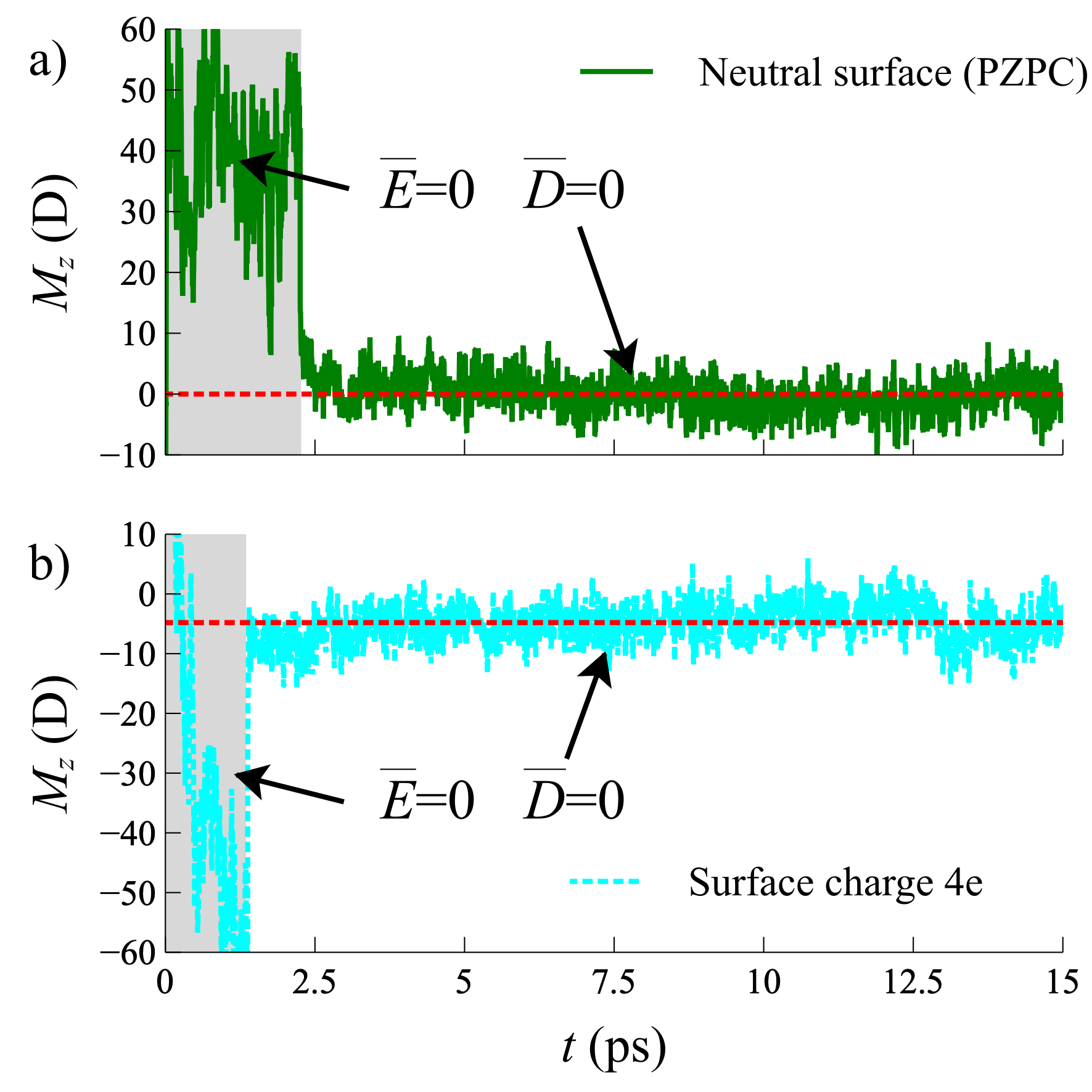}
\caption{\label{fig1} a) The time evolution of total dipole moment $M_z$ at PZPC when switching electric boundary condition from $\bar{E}=0$ to $\bar{D}=0$; b) The time evolution of total dipole moment $M_z$ with surface charge $4e$ when switching electric boundary condition from $\bar{E}=0$ to $\bar{D}=0$. Dash lines are the time average of $M_z$ in each case. }
\end{figure}

Conventionally, the Helmholtz capacitance can be computed as $C_\text{H}=q/\Delta\phi$ according to the textbook definition, where $q$ is the surface charge and $\Delta\phi$ is the potential drop crossing the Helmholtz layer~\cite{Limmer:2013ir,Cheng:2014eh}. Instead, by exploring the modern theory of polarization and constant $\bar{D}$ method developed by Vanderbilt and co-worker for treating ferroelectric nanocapacitors~\cite{Stengel:2009cd}, we reformulated this problem in terms of the macroscopic polarization $P_z$ of the system~\cite{Zhang:2016ca}. In our setup, the two sides of the oxide material can be charged at fixed chemical composition of the supercell by bearing same amount but opposite types of proton charges. This scheme leads to a succinct expression of the average Helmholtz capacitance $C_\text{H}$ based on the supercell polarization $P_z$~\cite{Zhang:2016ca}:

\begin{equation}
\label{C_H}
C_\text{H}=\left (\frac{\partial q}{2\pi L_z A \partial \langle P_z \rangle} \right )_{\bar{D}}
\end{equation}
where $L_z$ is the dimension of the model system perpendicular to the interface, $q$ is the imposed proton charge, $A$ is the area of the x, y cross section and $\langle \cdots \rangle$ indicates the ensemble average. The advantage of this formulation is three-fold. First, it does not require additional vacuum slab in the modeling as commonly used in surface science~\cite{Parez:2014dx,Creazzo:2019ek} and the oxide is treated as one piece of material. Second, it removes the finite-size dependence of the oxide slab which plagues the computation of the Helmholtz capacitance~\cite{Zhang:2016ca}. Third, by switching the electric boundary condition to constant $\bar{D}$, the relaxation of $P_z$ is significantly accelerated as predicted by the Debye theory of dielectrics~\cite{Zhang:2016cl,Zhang:2016ho}.

When the oxide surface is charged by protonation or deprotonation, $P_z$ deviates from zero in response. Therefore, $P_z$ has to be zero at the PZPC and this serves as a critical test for the convergence of DFTMD simulations~\cite{M_z}. As shown in Fig.~\ref{fig1}, at the PZPC, $M_z$ relaxes to zero rapidly when switching the boundary condition from $\bar{E}=0$ to $\bar{D}=0$. Within 10 picoseconds, its time average is about 0.02 D. For the case of electrified interface with surface charge of 4$e$, $M_z$ turns out to be -4.76 D within a similar time-scale. In all cases, classical MD simulations were leveraged to pre-equilibrate the system before applying the hybrid constant $\bar{D}$ DFTMD simulations~\cite{Zhang:2018cf, Sayer:2019bm}.

\begin{table}
\caption{\label{tab1} The averaged Helmholtz capacitance $C_\text{H}$, protonic Helmholtz capacitances $C_\text{H}^+$  and deprotonic Helmholtz capacitance $C_\text{H}^-$  at different surface charge $q$. For $q=2e$, the corresponding surface charge density for the supercell size used here is about 20 $\mu$C/cm$^2$.}
\begin{tabular*}{1.0\columnwidth}{ c c c c }
\hline
$q$ & $C_\text{H}$  ($\mu$F/cm$^2$) & $C_\text{H}^+$  ($\mu$F/cm$^2$)  & $C_\text{H}^-$  ($\mu$F/cm$^2$)\\
\hline
2$e$ &81 &67 &101 \\
  4$e$ & 72 & 59 & 85\\
\hline
\end{tabular*}
\end{table}

\begin{figure} [h]
\includegraphics[width=1.0\columnwidth]{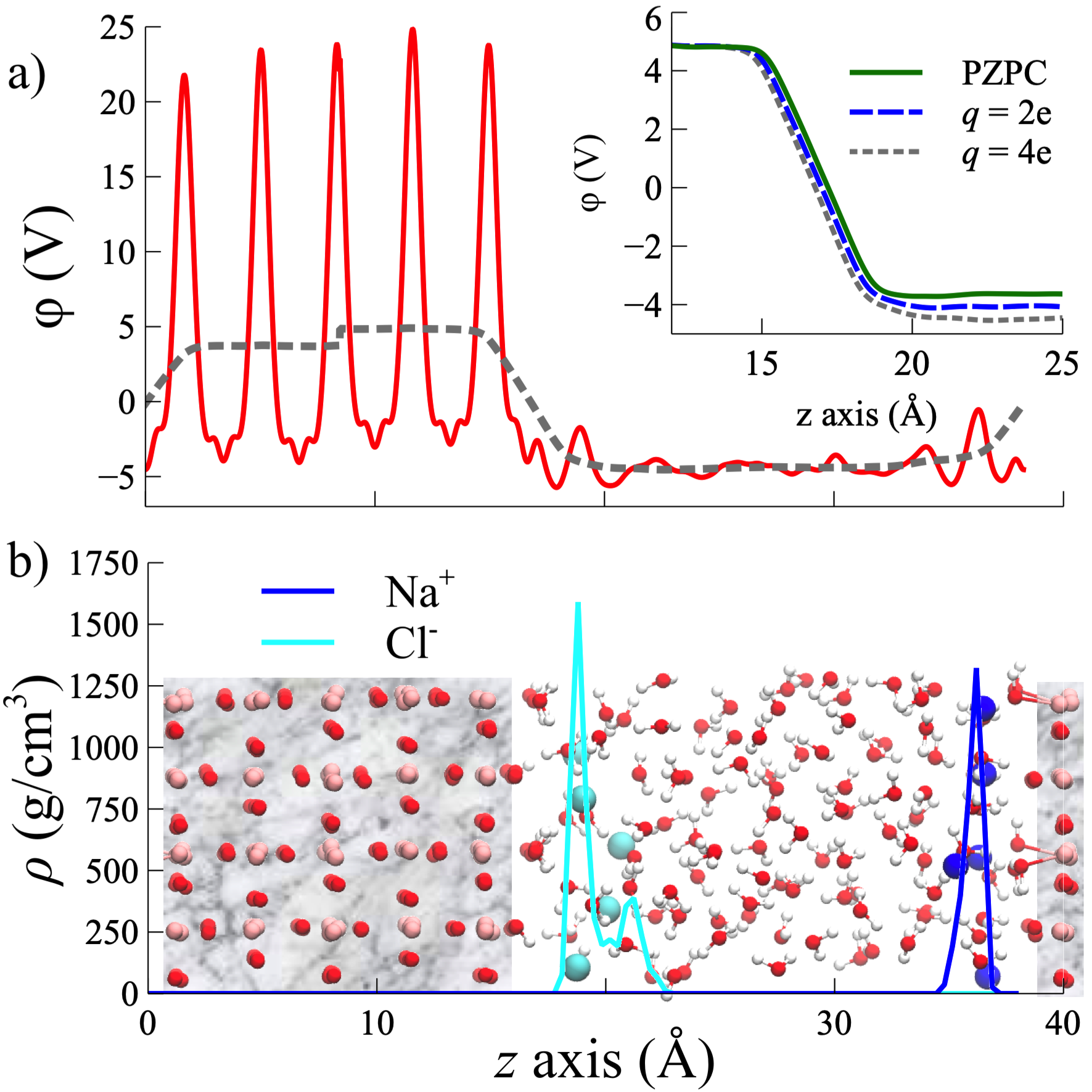}
\caption{\label{fig2} a) The planar averaged (solid line) and macro-averaged (dash line) electrostatic potential $\phi$ at surface charge $q=4e$. Inset: Macro-averaged electrostatic potential for the protonic  side at different surface charge densities; b) The corresponding density profiles of ions at surface charge $q=4e$ overlaid with one snapshot of the model system. The protonic side is formed by adding extra H$^+$ to the oxygen sites and the deprotonic side is formed by removing H$^+$ from the absorbed water molecules. Thus, the supercell composition is kept fixed at different surface charge densities.}
\end{figure}

Simulation results of the average Helmholtz capacitance $C_\text{H}$ at the rutile TiO$_2$ (110)-NaCl interface are shown in Table~\ref{tab1}. Following Eq.~\ref{C_H}, one obtains the average Helmholtz capacitance $C_\text{H}$ which is about 76 $\mu$F/cm$^2$. Although starting from very different initial configurations, $C_\text{H}$ at surface charge $q=2e$ and $q=4e$ are in excellence agreement with each other. In order to decompose the overall Helmholtz capacitance into protonic $C_\text{H}^+$ and deprotonic $C_\text{H}^-$, we resorted to the macroscopic averaging technique~\cite{Junquera:2007eb,si} and monitored the shift of the macro-averaged electrostatic potential with respect to the PZPC for the protonic side (Fig.~\ref{fig2} inset). From this decomposition, we found that $C_\text{H}^-$ is about 50\% higher than $C_\text{H}^+$. At surface charge $q=2e$ (about 20 $\mu$C/cm$^2$ for the supercell used here~\cite{si}), the surface potential for the deprotonic side is about 200 mV and that for the protonic side is about 300 mV, which can be measured in principle using the surface-sensitive vibrational spectroscopy~\cite{Ong:1992ca} or the binding energy shift in X-ray photoelectron spectroscopy~\cite{Brown:2016bo}.

The range of experimental estimated capacitance for rutile TiO$_2$-NaCl electrolyte goes from 64 to 160 $\mu$F/cm$^2$~\cite{Bourikas:2001hh, Ridley:2009ig, Berube:1968im}. The scattered data reflect the nature of the observed capacitance which depends on ionic strength and surface roughness~\cite{Hiemstra:1991ju}. In particular, the asymmetric pH-dependence (a higher Helmholtz capacitance at high pH) that we observed from DFTMD simulations is in accord with titration experiments of rutile at higher ionic strength~\cite{Berube:1968im, Yates:1980io, Kallay:1994iu} and has been also seen in other metal oxides, such as ZnO~\cite{Blok:1970jn}.

The Stern layer width (the charge separation distance) for the negatively charged TiO$_2$ surface is about 2\AA~ in our simulations (Fig.~\ref{fig2}b) and this leads to an estimation of the interfacial dielectric constant of about 23. This number can be compared with the commonly assumed value of 26 for rutile TiO$_2$ in geochemistry and colloid science~\cite{Sverjensky:2005fe}. On the other hand, we found the Stern layer width for the positively charge TiO$_2$ surface is about the same ($\sim$2\AA, Fig.~\ref{fig2}b), thus the smaller capacitance of the protonic side $C_\text{H}^+$ (Table~\ref{tab1}) suggests an interfacial dielectric constant of 15 instead. It is worth noting that the maxima in the radial distribution functions of Na$^+-$O$_\textrm{w}$ and Cl$^--$H$_\textrm{w}$ in bulk salt solutions are 2.4\AA~ and 2.9\AA~ respectively with the PBE functional~\cite{Bankura:2013bd}. Therefore, the asymmetry of the interfacial capacitances found here should largely come from the difference in the dielectric screening at the interface.

\begin{figure}
\includegraphics[width=1.0\columnwidth]{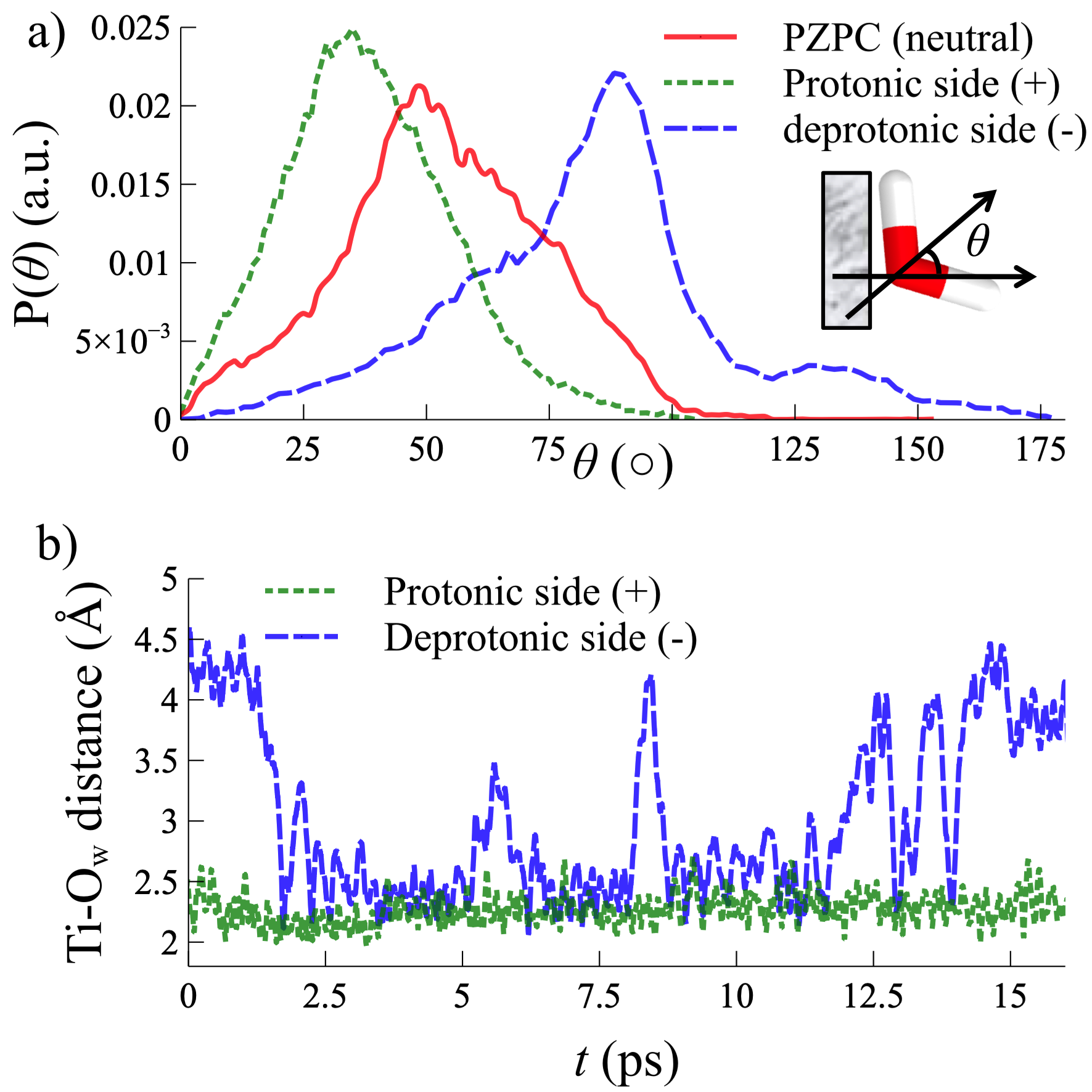}
\caption{\label{fig3} a) Probability distribution of the angle $\theta$ between water dipole moment and surface normal for neutral, positively and negatively charged ($q=2e$) rutile TiO$_2$ (110)-electrolyte interfaces ; b) Dynamics of adsorbed water molecules at positively and negatively charged ($q=2e$) rutile TiO$_2$ (110)-electrolyte interfaces.}
\end{figure}

At the PZPC condition, water molecules are adsorbed to the rutile TiO$_2$ (110) surface as  dimers~\cite{Serrano:2015it}. This is evidenced by the angular distribution of water dipole moments with respect to the normal vector of the surface (Fig.~\ref{fig3}a). Water dipole moment preferably points out towards the electrolyte solution with primary and secondary peaks at 48$^\circ$ and 75$^\circ$.  For the positively charged side, the electric field of EDL enhances this pattern by shifting primary peak and secondary peaks to 33$^\circ$ and 48$^\circ$ respectively. The splitting between the two peaks becomes narrower and the overall distribution is less broad in comparison with the neutral surface. The situation is reversed for the negatively charged side.  The angular distribution spans almost the whole range of possible values with primary and secondary peaks around 89$^\circ$ and 60 $^\circ$.

The much wider angular distribution of water molecules at the negatively charged side comes from two competing factors: chemisorption of water molecules at rutile (110) would orientate water dipole towards to the electrolyte while the electric field in EDL tends to flip it. Indeed, adsorption and desorption can happen dynamically at the negatively charged side in contrast to the positively charged side because of this competition, as seen in Fig.~\ref{fig3}b.  Since the extent of dielectric screening is proportional to the magnitude of dipole moment fluctuation as formulated in Kirkwood-Onsager theory~\cite{Kirkwood:1939br, Zhang:2016ho}, it is not a surprise that we found the interfacial dielectric constant at the deprotonic side (therefore $C_{\text{H}}^-$) is much higher than that of protonic one.

This dynamical adsorption-desorption process of water molecules observed at the negatively charged side may play a role for the alternative mechanism of O$_2$ production in alkaline solution (high pH)~\cite{Imanishi:2014cj}, since adsorbed water molecules at the TiO$_2$ surface not only provide the raw material for OH$\cdot$ production but may also block surface sites~\cite{Henderson:2011id}.

\begin{figure}
\includegraphics[width=1.0\columnwidth]{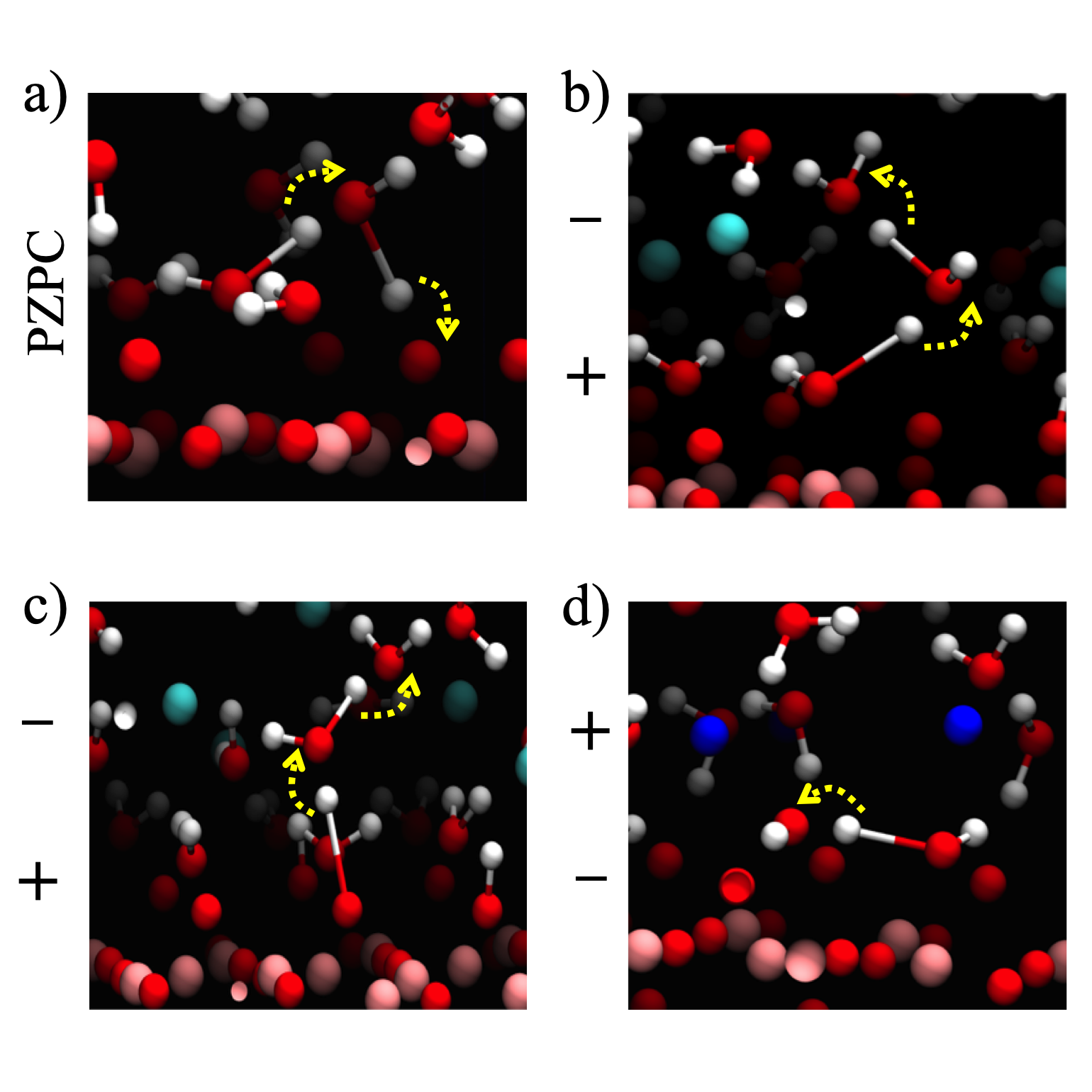}
\caption{\label{fig4} a) Dissociative adsorption of water molecules at the neutral surface (PZPC); b) Hydrolysis of adsorbed water molecules at the positively charged surface; c) Proton transfer between charged surface groups Ti$_2$OH$^+$ and solvating water molecules at the positively charged surface. Ti$_2$O indicates the bridging oxygens; d) Resonance between charged surface groups TiOH$^-$  and adsorbed water molecules at the negatively charged side. TiOH$^-$ stand for the hydroxylated 5-fold coordinated Ti groups. }
\end{figure}

The dissociative adsorption of water molecules at the bare rutile TiO$_2$ (110) surface was extensively discussed in literature~\cite{Liu:2010co, Anonymous:2012dr, Liu:2012bd}. Indeed, this is also observed for the neutral surface mediated by the neighboring water molecule (Ref.~\cite{Tocci:2014by} and Fig.~\ref{fig4} a). One would expect the electric field due to the EDL will magnify this effect. However, instead of the dissociative adsorption, the hydrolysis of adsorbed water molecules releases ions to the electrolyte solution. In our simulations, this dominantly happens for the positively charged rutile TiO$_2$ (110) surface (Fig.~\ref{fig4} b), in accord with a much higher electric field generated in the protonic side.

The pKa of the protonic rutile TiO$_2$ (110) was determined to be about -1.0 in DFTMD~\cite{Cheng:2010gg}. Therefore,  the estimated reaction free energy for the proton transfer between  the charged surface groups Ti$_2$OH$^+$ and solvating water molecules is comparable to thermal fluctuation energy at room temperature.  Indeed, spontaneous proton exchange between them was observed for the protonic side of rutile TiO$_2$ (110), as shown in Fig.~\ref{fig4} c. On the contrary, the exchange of OH$^-$ between TiOH$^-$ and solvating water molecules is endothermic and no corresponding OH$^-$ transfer event was observed at the deprotonic side. Instead, the resonance of charged group TiOH$^-$ and surface groups with adsorbed water molecules TiH$_2$O happens (Fig.~\ref{fig4} d). 

To quantify the effect of surface acid-base chemistry on the Helmholtz capacitance, we also carried out additional simulations by constraining adsorbed water molecules and charged surface groups Ti$_2$OH$^+$ to not undergo reaction. When only constraining adsorbed water molecules, the average Helmholtz capacitance $C_\text{H}$ at $q=4e$ turns out to be about 80$\mu$F/cm$^2$. This value is very similar to the one about 72$\mu$F/cm$^2$ in Table~\ref{tab1} without constraints. However, when both adsorbed water molecules and charged surface groups are constrained, $C_\text{H}$ becomes about 44$\mu$F/cm$^2$ which is close to results obtained previously~\cite{Cheng:2014eh}. This significant reduction of capacitance can be understood, because the proton exchange between Ti$_2$OH$^+$ and solvating water molecules will shorten the charge separation distance in the EDL. Therefore, the surface acidity of the oxide is another determining factor for the Helmholtz capacitance.

In summary, using finite-field DFTMD simulations, we discovered the microscopic origin for the pH-dependent of Helmholtz capacitance at rutile TiO$_2$ (110) seen in titration experiments. At high pH, water molecules have a stronger structural fluctuation and this lead to a much larger capacitance. At low pH, proton transfer increases the capacitance value by reducing the charge seperation distance. These observations for rutile TiO$_2$ (110) and the finite-field DFTMD modeling approach presented in this study pave the way to investigate electrochemical reactivity at electrified metal oxide-electrolyte interfaces, e.g. the Non-Nernstian contribution of the capacitance to the Pourbaix diagram and the hole trapping in alkaline solutions. 

\begin{acknowledgement}
CZ thanks Deutsche Forschungsgemeinschaft (DFG) for a research fellowship (No.~ZH 477/1-1) during his stay in Cambridge. CZ is grateful to Uppsala University for a start-up grant and to \AA forsk Foundation for a research grant (Ref. nr. 18-460). Funding from the Swedish National Strategic e-Science program eSSENCE is also gratefully acknowledged. Computational resources were provided by the UK Car-Parrinello (UKCP) consortium funded by the Engineering and Physical Sciences Research Council (EPSRC) of the United Kingdom.  CZ also thanks J. Cheng for helpful discussions and T. Sayer for reading the manuscript.
\end{acknowledgement}

\begin{suppinfo}

The hybrid constant $\bar{D}$ Hamiltonian and its corresponding CP2K input syntax. The details of CP2K simulations of electrified rutile TiO$_2$ (110)-NaCl electrolyte interfaces.

\end{suppinfo}


\begin{mcitethebibliography}{78}
\providecommand*\natexlab[1]{#1}
\providecommand*\mciteSetBstSublistMode[1]{}
\providecommand*\mciteSetBstMaxWidthForm[2]{}
\providecommand*\mciteBstWouldAddEndPuncttrue
  {\def\EndOfBibitem{\unskip.}}
\providecommand*\mciteBstWouldAddEndPunctfalse
  {\let\EndOfBibitem\relax}
\providecommand*\mciteSetBstMidEndSepPunct[3]{}
\providecommand*\mciteSetBstSublistLabelBeginEnd[3]{}
\providecommand*\EndOfBibitem{}
\mciteSetBstSublistMode{f}
\mciteSetBstMaxWidthForm{subitem}{(\alph{mcitesubitemcount})}
\mciteSetBstSublistLabelBeginEnd
  {\mcitemaxwidthsubitemform\space}
  {\relax}
  {\relax}

\bibitem[Ardizzone and Trasatti(1996)Ardizzone, and Trasatti]{Ardizzone:1996ca}
Ardizzone,~S.; Trasatti,~S. {Interfacial properties of oxides with
  technological impact in electrochemistry}. \emph{Adv. Colloid Interface Sci.}
  \textbf{1996}, \emph{64}, 173--251\relax
\mciteBstWouldAddEndPuncttrue
\mciteSetBstMidEndSepPunct{\mcitedefaultmidpunct}
{\mcitedefaultendpunct}{\mcitedefaultseppunct}\relax
\EndOfBibitem
\bibitem[Gross and Sakong(2019)Gross, and Sakong]{GRO20191}
Gross,~A.; Sakong,~S. Modelling the electric double layer at
  electrode/electrolyte interfaces. \emph{Curr. Opin. Electrochem.}
  \textbf{2019}, \emph{14}, 1 -- 6\relax
\mciteBstWouldAddEndPuncttrue
\mciteSetBstMidEndSepPunct{\mcitedefaultmidpunct}
{\mcitedefaultendpunct}{\mcitedefaultseppunct}\relax
\EndOfBibitem
\bibitem[Adekola \latin{et~al.}(2011)Adekola, F{\'e}doroff, Geckeis, Kupcik,
  Lef{\`e}vre, L{\"u}tzenkirchen, Plaschke, Preo{\v c}anin, Rabung, and
  Schild]{Adekola:2011kk}
Adekola,~F.; F{\'e}doroff,~M.; Geckeis,~H.; Kupcik,~T.; Lef{\`e}vre,~G.;
  L{\"u}tzenkirchen,~J.; Plaschke,~M.; Preo{\v c}anin,~T.; Rabung,~T.;
  Schild,~D. {Characterization of acid-base properties of two gibbsite samples
  in the context of literature results}. \emph{J. Colloid Interface Sci.}
  \textbf{2011}, \emph{354}, 306--317\relax
\mciteBstWouldAddEndPuncttrue
\mciteSetBstMidEndSepPunct{\mcitedefaultmidpunct}
{\mcitedefaultendpunct}{\mcitedefaultseppunct}\relax
\EndOfBibitem
\bibitem[Zarzycki and Preo{\v c}anin(2012)Zarzycki, and Preo{\v
  c}anin]{Zarzycki:2012ds}
Zarzycki,~P.; Preo{\v c}anin,~T. {Point of zero potential of single-crystal
  electrode/inert electrolyte interface}. \emph{J. Colloid Interface Sci.}
  \textbf{2012}, \emph{370}, 139--143\relax
\mciteBstWouldAddEndPuncttrue
\mciteSetBstMidEndSepPunct{\mcitedefaultmidpunct}
{\mcitedefaultendpunct}{\mcitedefaultseppunct}\relax
\EndOfBibitem
\bibitem[Kosmulski(2014)]{Kosmulski:2014fv}
Kosmulski,~M. {The pH dependent surface charging and points of zero charge. VI.
  Update}. \emph{J. Colloid Interface Sci.} \textbf{2014}, \emph{426},
  209--212\relax
\mciteBstWouldAddEndPuncttrue
\mciteSetBstMidEndSepPunct{\mcitedefaultmidpunct}
{\mcitedefaultendpunct}{\mcitedefaultseppunct}\relax
\EndOfBibitem
\bibitem[Lyons \latin{et~al.}(2017)Lyons, Doyle, Browne, Godwin, and
  Rovetta]{Lyons:2017cg}
Lyons,~M. E.~G.; Doyle,~R.~L.; Browne,~M.~P.; Godwin,~I.~J.; Rovetta,~A. A.~S.
  {Recent developments in electrochemical water oxidation}.
  \emph{Curr. Opin. Electrochem.} \textbf{2017}, \emph{1}, 40--45\relax
\mciteBstWouldAddEndPuncttrue
\mciteSetBstMidEndSepPunct{\mcitedefaultmidpunct}
{\mcitedefaultendpunct}{\mcitedefaultseppunct}\relax
\EndOfBibitem
\bibitem[Schmickler and Santos(2010)Schmickler, and Santos]{Schmickler:2010th}
Schmickler,~W.; Santos,~E. \emph{{Interfacial electrochemistry}}; Springer:
  Berlin; London, 2010\relax
\mciteBstWouldAddEndPuncttrue
\mciteSetBstMidEndSepPunct{\mcitedefaultmidpunct}
{\mcitedefaultendpunct}{\mcitedefaultseppunct}\relax
\EndOfBibitem
\bibitem[Nozik and Memming(1996)Nozik, and Memming]{Nozik:1996gv}
Nozik,~A.~J.; Memming,~R. {Physical chemistry of semiconductor-liquid
  interfaces}. \emph{J. Phys. Chem.} \textbf{1996}, \emph{100},
  13061--13078\relax
\mciteBstWouldAddEndPuncttrue
\mciteSetBstMidEndSepPunct{\mcitedefaultmidpunct}
{\mcitedefaultendpunct}{\mcitedefaultseppunct}\relax
\EndOfBibitem
\bibitem[Gr{\"a}tzel(2001)]{Gratzel:2001ub}
Gr{\"a}tzel,~M. {Photoelectrochemical cells}. \emph{Nature} \textbf{2001},
  \emph{414}, 338--344\relax
\mciteBstWouldAddEndPuncttrue
\mciteSetBstMidEndSepPunct{\mcitedefaultmidpunct}
{\mcitedefaultendpunct}{\mcitedefaultseppunct}\relax
\EndOfBibitem
\bibitem[Bolt(1957)]{BOLT:1957dr}
Bolt,~G.~H. {Determination of the Charge Density of Silica Sols}. \emph{J.
  Phys. Chem.} \textbf{1957}, \emph{61}, 1166--1169\relax
\mciteBstWouldAddEndPuncttrue
\mciteSetBstMidEndSepPunct{\mcitedefaultmidpunct}
{\mcitedefaultendpunct}{\mcitedefaultseppunct}\relax
\EndOfBibitem
\bibitem[Parks and de~Bruyn(1962)Parks, and de~Bruyn]{Parks:1962eva}
Parks,~G.~A.; de~Bruyn,~P.~L. {The zero point of charge of oxides}. \emph{J.
  Phys. Chem.} \textbf{1962}, \emph{66}, 967--973\relax
\mciteBstWouldAddEndPuncttrue
\mciteSetBstMidEndSepPunct{\mcitedefaultmidpunct}
{\mcitedefaultendpunct}{\mcitedefaultseppunct}\relax
\EndOfBibitem
\bibitem[Parks(1965)]{Parks:1965gaa}
Parks,~G.~A. {The Isoelectric Points of Solid Oxides, Solid Hydroxides, and
  Aqueous Hydroxo Complex Systems}. \emph{Chem. Rev.} \textbf{1965}, \emph{65},
  177--198\relax
\mciteBstWouldAddEndPuncttrue
\mciteSetBstMidEndSepPunct{\mcitedefaultmidpunct}
{\mcitedefaultendpunct}{\mcitedefaultseppunct}\relax
\EndOfBibitem
\bibitem[Hiemstra \latin{et~al.}(1989)Hiemstra, van Riemsdijk, and
  Bolt]{Hiemstra:1989vh}
Hiemstra,~T.; van Riemsdijk,~W.~H.; Bolt,~G.~H. {Multisite Proton Adsorption
  Modeling at the Solid-Solution Interface of (Hydr)Oxides - a New Approach .1.
  Model Description and Evaluation of Intrinsic Reaction Constants}. \emph{J.
  Colloid Interface Sci.} \textbf{1989}, \emph{133}, 91--104\relax
\mciteBstWouldAddEndPuncttrue
\mciteSetBstMidEndSepPunct{\mcitedefaultmidpunct}
{\mcitedefaultendpunct}{\mcitedefaultseppunct}\relax
\EndOfBibitem
\bibitem[Hiemstra \latin{et~al.}(1989)Hiemstra, De~Wit, and
  Van~Riemsdijk]{Hiemstra:1989br}
Hiemstra,~T.; De~Wit,~J. C.~M.; Van~Riemsdijk,~W.~H. {Multisite proton
  adsorption modeling at the solid/solution interface of (hydr)oxides: A new
  approach. II. Application to various important (hydr)oxides}. \emph{J.
  Colloid Interface Sci.} \textbf{1989}, \emph{133}, --117\relax
\mciteBstWouldAddEndPuncttrue
\mciteSetBstMidEndSepPunct{\mcitedefaultmidpunct}
{\mcitedefaultendpunct}{\mcitedefaultseppunct}\relax
\EndOfBibitem
\bibitem[Bickmore \latin{et~al.}(2004)Bickmore, Tadanier, Rosso, Monn, and
  Eggett]{Bickmore:2004gya}
Bickmore,~B.~R.; Tadanier,~C.~J.; Rosso,~K.~M.; Monn,~W.~D.; Eggett,~D.~L.
  {Bond-valence methods for pKa prediction: critical reanalysis and a new
  approach}. \emph{Geochim. Cosmochim. Acta} \textbf{2004}, \emph{68},
  2025--2042\relax
\mciteBstWouldAddEndPuncttrue
\mciteSetBstMidEndSepPunct{\mcitedefaultmidpunct}
{\mcitedefaultendpunct}{\mcitedefaultseppunct}\relax
\EndOfBibitem
\bibitem[Brown(2009)]{Brown:2009iu}
Brown,~I.~D. {Recent Developments in the Methods and Applications of the Bond
  Valence Model}. \emph{Chem. Rev.} \textbf{2009}, \emph{109}, 6858--6919\relax
\mciteBstWouldAddEndPuncttrue
\mciteSetBstMidEndSepPunct{\mcitedefaultmidpunct}
{\mcitedefaultendpunct}{\mcitedefaultseppunct}\relax
\EndOfBibitem
\bibitem[Bickmore \latin{et~al.}(2006)Bickmore, Rosso, and
  Mitchell]{BICKMORE2006269}
Bickmore,~B.; Rosso,~K.; Mitchell,~S. In \emph{Surface Complexation Modelling};
  Lützenkirchen,~J., Ed.; Interface Science and Technology; Elsevier, 2006;
  Vol.~11; pp 269 -- 283\relax
\mciteBstWouldAddEndPuncttrue
\mciteSetBstMidEndSepPunct{\mcitedefaultmidpunct}
{\mcitedefaultendpunct}{\mcitedefaultseppunct}\relax
\EndOfBibitem
\bibitem[Boily(2014)]{Boily:2014jp}
Boily,~J.-F. {The Variable Capacitance Model: A Strategy for Treating
  Contrasting Charge-Neutralizing Capabilities of Counterions at the
  Mineral/Water Interface}. \emph{Langmuir} \textbf{2014}, \emph{30},
  2009--2018\relax
\mciteBstWouldAddEndPuncttrue
\mciteSetBstMidEndSepPunct{\mcitedefaultmidpunct}
{\mcitedefaultendpunct}{\mcitedefaultseppunct}\relax
\EndOfBibitem
\bibitem[Boamah \latin{et~al.}(2018)Boamah, Ohno, Geiger, and
  Eisenthal]{Boamah:2018kn}
Boamah,~M.~D.; Ohno,~P.~E.; Geiger,~F.~M.; Eisenthal,~K.~B. {Relative
  permittivity in the electrical double layer from nonlinear optics}. \emph{J.
  Chem. Phys.} \textbf{2018}, \emph{148}, 222808--8\relax
\mciteBstWouldAddEndPuncttrue
\mciteSetBstMidEndSepPunct{\mcitedefaultmidpunct}
{\mcitedefaultendpunct}{\mcitedefaultseppunct}\relax
\EndOfBibitem
\bibitem[Fujishima and Honda(1972)Fujishima, and Honda]{FUJISHIMA:1972hc}
Fujishima,~A.; Honda,~K. {Electrochemical Photolysis of Water at a
  Semiconductor Electrode}. \emph{Nature} \textbf{1972}, \emph{238},
  37--38\relax
\mciteBstWouldAddEndPuncttrue
\mciteSetBstMidEndSepPunct{\mcitedefaultmidpunct}
{\mcitedefaultendpunct}{\mcitedefaultseppunct}\relax
\EndOfBibitem
\bibitem[Gerischer(1978)]{Gerischer:1978ig}
Gerischer,~H. {Electrolytic decomposition and photodecomposition of compound
  semiconductors in contact with electrolytes}. \emph{J. Vac. Sci. Technol.}
  \textbf{1978}, \emph{15}, 1422--1428\relax
\mciteBstWouldAddEndPuncttrue
\mciteSetBstMidEndSepPunct{\mcitedefaultmidpunct}
{\mcitedefaultendpunct}{\mcitedefaultseppunct}\relax
\EndOfBibitem
\bibitem[Henderson(2011)]{Henderson:2011id}
Henderson,~M.~A. {A surface science perspective on TiO2 photocatalysis}.
  \emph{Surf. Sci. Rep.} \textbf{2011}, \emph{66}, 185--297\relax
\mciteBstWouldAddEndPuncttrue
\mciteSetBstMidEndSepPunct{\mcitedefaultmidpunct}
{\mcitedefaultendpunct}{\mcitedefaultseppunct}\relax
\EndOfBibitem
\bibitem[Pang \latin{et~al.}(2013)Pang, Lindsay, and Thornton]{Pang:2013fwa}
Pang,~C.~L.; Lindsay,~R.; Thornton,~G. {Structure of Clean and
  Adsorbate-Covered Single-Crystal Rutile TiO2 Surfaces}. \emph{Chem. Rev.}
  \textbf{2013}, \emph{113}, 3887--3948\relax
\mciteBstWouldAddEndPuncttrue
\mciteSetBstMidEndSepPunct{\mcitedefaultmidpunct}
{\mcitedefaultendpunct}{\mcitedefaultseppunct}\relax
\EndOfBibitem
\bibitem[Hussain \latin{et~al.}(2016)Hussain, Tocci, Woolcot, Torrelles, Pang,
  Humphrey, Yim, Grinter, Cabailh, Bikondoa, Lindsay, Zegenhagen, Michaelides,
  and Thornton]{Hussain:2016bu}
Hussain,~H.; Tocci,~G.; Woolcot,~T.; Torrelles,~X.; Pang,~C.~L.;
  Humphrey,~D.~S.; Yim,~C.~M.; Grinter,~D.~C.; Cabailh,~G.; Bikondoa,~O.
  \latin{et~al.}  {Structure of a model TiO2~photocatalytic interface}.
  \emph{Nat. Mater.} \textbf{2016}, 1--7\relax
\mciteBstWouldAddEndPuncttrue
\mciteSetBstMidEndSepPunct{\mcitedefaultmidpunct}
{\mcitedefaultendpunct}{\mcitedefaultseppunct}\relax
\EndOfBibitem
\bibitem[Diebold(2017)]{Diebold:2017bp}
Diebold,~U. {Perspective: A controversial benchmark system for water-oxide
  interfaces: H2O/TiO2(110)}. \emph{J. Chem. Phys.} \textbf{2017}, \emph{147},
  040901--4\relax
\mciteBstWouldAddEndPuncttrue
\mciteSetBstMidEndSepPunct{\mcitedefaultmidpunct}
{\mcitedefaultendpunct}{\mcitedefaultseppunct}\relax
\EndOfBibitem
\bibitem[Balajka \latin{et~al.}(2018)Balajka, Pavelec, Komora, Schmid, and
  Diebold]{Balajka:2018gw}
Balajka,~J.; Pavelec,~J.; Komora,~M.; Schmid,~M.; Diebold,~U. {Apparatus for
  dosing liquid water in ultrahigh vacuum}. \emph{Rev. Sci. Instrum.}
  \textbf{2018}, \emph{89}, 083906--7\relax
\mciteBstWouldAddEndPuncttrue
\mciteSetBstMidEndSepPunct{\mcitedefaultmidpunct}
{\mcitedefaultendpunct}{\mcitedefaultseppunct}\relax
\EndOfBibitem
\bibitem[Tomkiewicz(1979)]{Tomkiewicz:1979ifa}
Tomkiewicz,~M. {The Potential Distribution at the TiO2 Aqueous Electrolyte
  Interface}. \emph{J. Electrochem. Soc.} \textbf{1979}, \emph{126},
  1505--1510\relax
\mciteBstWouldAddEndPuncttrue
\mciteSetBstMidEndSepPunct{\mcitedefaultmidpunct}
{\mcitedefaultendpunct}{\mcitedefaultseppunct}\relax
\EndOfBibitem
\bibitem[Ridley \latin{et~al.}(2009)Ridley, Hiemstra, van Riemsdijk, and
  Machesky]{Ridley:2009ig}
Ridley,~M.~K.; Hiemstra,~T.; van Riemsdijk,~W.~H.; Machesky,~M.~L.
  {Inner-sphere complexation of cations at the rutile{\textendash}water
  interface: A concise surface structural interpretation with the CD and MUSIC
  model}. \emph{Geochim. Cosmochim. Acta} \textbf{2009}, \emph{73},
  1841--1856\relax
\mciteBstWouldAddEndPuncttrue
\mciteSetBstMidEndSepPunct{\mcitedefaultmidpunct}
{\mcitedefaultendpunct}{\mcitedefaultseppunct}\relax
\EndOfBibitem
\bibitem[B{\'e}rub{\'e} and de~Bruyn(1968)B{\'e}rub{\'e}, and
  de~Bruyn]{Berube:1968im}
B{\'e}rub{\'e},~Y.~G.; de~Bruyn,~P.~L. {Adsorption at the rutile-solution
  interface. II. Model of the electrochemical double layer}. \emph{J. Colloid
  Interface Sci.} \textbf{1968}, \emph{28}, 92--105\relax
\mciteBstWouldAddEndPuncttrue
\mciteSetBstMidEndSepPunct{\mcitedefaultmidpunct}
{\mcitedefaultendpunct}{\mcitedefaultseppunct}\relax
\EndOfBibitem
\bibitem[Gerischer(1989)]{GERISCHER19891005}
Gerischer,~H. Neglected problems in the pH dependence of the flatband potential
  of semiconducting\ oxides and semiconductors covered with oxide layers.
  \emph{Electrochim. Acta} \textbf{1989}, \emph{34}, 1005 -- 1009\relax
\mciteBstWouldAddEndPuncttrue
\mciteSetBstMidEndSepPunct{\mcitedefaultmidpunct}
{\mcitedefaultendpunct}{\mcitedefaultseppunct}\relax
\EndOfBibitem
\bibitem[Gileadi(2011)]{Gileadi:2011fca}
Gileadi,~E. {Problems in interfacial electrochemistry that have been swept
  under the carpet}. \emph{J. Solid State Electrochem.} \textbf{2011},
  \emph{15}, 1359--1371\relax
\mciteBstWouldAddEndPuncttrue
\mciteSetBstMidEndSepPunct{\mcitedefaultmidpunct}
{\mcitedefaultendpunct}{\mcitedefaultseppunct}\relax
\EndOfBibitem
\bibitem[Liu \latin{et~al.}(2010)Liu, Zhang, Thornton, and
  Michaelides]{Liu:2010co}
Liu,~L.-M.; Zhang,~C.; Thornton,~G.; Michaelides,~A. {Structure and dynamics of
  liquid water on rutile TiO2(110)}. \emph{Phys. Rev. B} \textbf{2010},
  \emph{82}, 161415--4\relax
\mciteBstWouldAddEndPuncttrue
\mciteSetBstMidEndSepPunct{\mcitedefaultmidpunct}
{\mcitedefaultendpunct}{\mcitedefaultseppunct}\relax
\EndOfBibitem
\bibitem[Wesolowski \latin{et~al.}(2012)Wesolowski, Sofo, Bandura, Zhang,
  Mamontov, P{\v{r}}edota, Kumar, Kubicki, Kent, Vlcek, Machesky, Fenter,
  Cummings, Anovitz, Skelton, and Rosenqvist]{Anonymous:2012dr}
Wesolowski,~D.~J.; Sofo,~J.~O.; Bandura,~A.~V.; Zhang,~Z.; Mamontov,~E.;
  P{\v{r}}edota,~M.; Kumar,~N.; Kubicki,~J.~D.; Kent,~P. R.~C.; Vlcek,~L.
  \latin{et~al.}  {Comment on "Structure and dynamics of liquid water on rutile
  TiO2(110)"}. \emph{Phys. Rev. B} \textbf{2012}, \emph{85}, 167401\relax
\mciteBstWouldAddEndPuncttrue
\mciteSetBstMidEndSepPunct{\mcitedefaultmidpunct}
{\mcitedefaultendpunct}{\mcitedefaultseppunct}\relax
\EndOfBibitem
\bibitem[Liu \latin{et~al.}(2012)Liu, Zhang, Thornton, and
  Michaelides]{Liu:2012bd}
Liu,~L.-M.; Zhang,~C.; Thornton,~G.; Michaelides,~A. {Reply to Comment on
  `Structure and dynamics of liquid water on rutile TiO2 (110)'}. \emph{Phys.
  Rev. B} \textbf{2012}, \emph{85}, 167402\relax
\mciteBstWouldAddEndPuncttrue
\mciteSetBstMidEndSepPunct{\mcitedefaultmidpunct}
{\mcitedefaultendpunct}{\mcitedefaultseppunct}\relax
\EndOfBibitem
\bibitem[Bandura \latin{et~al.}(2011)Bandura, Kubicki, and
  Sofo]{Bandura:2011fg}
Bandura,~A.~V.; Kubicki,~J.~D.; Sofo,~J.~O. {Periodic Density Functional Theory
  Study of Water Adsorption on the $\alpha$-Quartz (101) Surface}. \emph{J.
  Phys. Chem. C} \textbf{2011}, \emph{115}, 5756--5766\relax
\mciteBstWouldAddEndPuncttrue
\mciteSetBstMidEndSepPunct{\mcitedefaultmidpunct}
{\mcitedefaultendpunct}{\mcitedefaultseppunct}\relax
\EndOfBibitem
\bibitem[Tocci and Michaelides(2014)Tocci, and Michaelides]{Tocci:2014by}
Tocci,~G.; Michaelides,~A. {Solvent-Induced Proton Hopping at a
  Water{\textendash}Oxide Interface}. \emph{J. Phys. Chem. Lett.}
  \textbf{2014}, \emph{5}, 474--480\relax
\mciteBstWouldAddEndPuncttrue
\mciteSetBstMidEndSepPunct{\mcitedefaultmidpunct}
{\mcitedefaultendpunct}{\mcitedefaultseppunct}\relax
\EndOfBibitem
\bibitem[von Rudorff \latin{et~al.}(2016)von Rudorff, Jakobsen, Rosso, and
  Blumberger]{vonRudorff:2016cf}
von Rudorff,~G.~F.; Jakobsen,~R.; Rosso,~K.~M.; Blumberger,~J. {Fast
  Interconversion of Hydrogen Bonding at the Hematite (001){\textendash}Liquid
  Water Interface}. \emph{J. Phys. Chem. Lett.} \textbf{2016}, \emph{7},
  1155--1160\relax
\mciteBstWouldAddEndPuncttrue
\mciteSetBstMidEndSepPunct{\mcitedefaultmidpunct}
{\mcitedefaultendpunct}{\mcitedefaultseppunct}\relax
\EndOfBibitem
\bibitem[Selcuk and Selloni(2016)Selcuk, and Selloni]{Selcuk:2016cb}
Selcuk,~S.; Selloni,~A. {Facet-dependent trapping and dynamics of excess
  electrons at anatase TiO2 surfaces and aqueous interfaces}. \emph{Nat.
  Mater.} \textbf{2016}, \emph{15}, 1107--1112\relax
\mciteBstWouldAddEndPuncttrue
\mciteSetBstMidEndSepPunct{\mcitedefaultmidpunct}
{\mcitedefaultendpunct}{\mcitedefaultseppunct}\relax
\EndOfBibitem
\bibitem[Gaigeot \latin{et~al.}(2012)Gaigeot, Sprik, and
  Sulpizi]{Anonymous:2012fp}
Gaigeot,~M.-P.; Sprik,~M.; Sulpizi,~M. {Oxide/water interfaces: how the surface
  chemistry modifies interfacial water properties}. \emph{J. Phys-Condens.
  Mat.} \textbf{2012}, \emph{24}, 124106--11\relax
\mciteBstWouldAddEndPuncttrue
\mciteSetBstMidEndSepPunct{\mcitedefaultmidpunct}
{\mcitedefaultendpunct}{\mcitedefaultseppunct}\relax
\EndOfBibitem
\bibitem[Wan and Galli(2015)Wan, and Galli]{Wan:2015id}
Wan,~Q.; Galli,~G. {First-Principles Framework to Compute Sum-Frequency
  Generation Vibrational Spectra of Semiconductors and Insulators}. \emph{Phys.
  Rev. Lett.} \textbf{2015}, \emph{115}, 246404--5\relax
\mciteBstWouldAddEndPuncttrue
\mciteSetBstMidEndSepPunct{\mcitedefaultmidpunct}
{\mcitedefaultendpunct}{\mcitedefaultseppunct}\relax
\EndOfBibitem
\bibitem[Khatib \latin{et~al.}(2016)Khatib, Backus, Bonn, Perez-Haro, Gaigeot,
  and Sulpizi]{Anonymous:2016be}
Khatib,~R.; Backus,~E. H.~G.; Bonn,~M.; Perez-Haro,~M.-J.; Gaigeot,~M.-P.;
  Sulpizi,~M. {Water orientation and hydrogen- bond structure at the
  fluorite/water interface}. \emph{Sci. Rep.} \textbf{2016}, \emph{6},
  24287\relax
\mciteBstWouldAddEndPuncttrue
\mciteSetBstMidEndSepPunct{\mcitedefaultmidpunct}
{\mcitedefaultendpunct}{\mcitedefaultseppunct}\relax
\EndOfBibitem
\bibitem[Cheng and Sprik(2010)Cheng, and Sprik]{Cheng:2010gg}
Cheng,~J.; Sprik,~M. {Acidity of the Aqueous Rutile TiO2(110) Surface from
  Density Functional Theory Based Molecular Dynamics}. \emph{J. Chem. Theory
  Comput.} \textbf{2010}, \emph{6}, 880--889\relax
\mciteBstWouldAddEndPuncttrue
\mciteSetBstMidEndSepPunct{\mcitedefaultmidpunct}
{\mcitedefaultendpunct}{\mcitedefaultseppunct}\relax
\EndOfBibitem
\bibitem[Sulpizi \latin{et~al.}(2012)Sulpizi, Gaigeot, and
  Sprik]{Sulpizi:2012ti}
Sulpizi,~M.; Gaigeot,~M.-P.; Sprik,~M. {The Silica{\textendash}Water Interface:
  How the Silanols Determine the Surface Acidity and Modulate the Water
  Properties}. \emph{J. Chem. Theory Comput.} \textbf{2012}, \emph{8},
  1037--1047\relax
\mciteBstWouldAddEndPuncttrue
\mciteSetBstMidEndSepPunct{\mcitedefaultmidpunct}
{\mcitedefaultendpunct}{\mcitedefaultseppunct}\relax
\EndOfBibitem
\bibitem[Liu \latin{et~al.}(2013)Liu, Lu, Sprik, Cheng, Meijer, and
  Wang]{Liu:2013he}
Liu,~X.; Lu,~X.; Sprik,~M.; Cheng,~J.; Meijer,~E.~J.; Wang,~R. {Acidity of edge
  surface sites of montmorillonite and kaolinite}. \emph{Geochim. Cosmochim.
  Acta} \textbf{2013}, \emph{117}, 180--190\relax
\mciteBstWouldAddEndPuncttrue
\mciteSetBstMidEndSepPunct{\mcitedefaultmidpunct}
{\mcitedefaultendpunct}{\mcitedefaultseppunct}\relax
\EndOfBibitem
\bibitem[Churakov \latin{et~al.}(2014)Churakov, Labbez, Pegado, and
  Sulpizi]{Churakov:2014bl}
Churakov,~S.~V.; Labbez,~C.; Pegado,~L.; Sulpizi,~M. {Intrinsic Acidity of
  Surface Sites in Calcium Silicate Hydrates and Its Implication to Their
  Electrokinetic Properties}. \emph{J. Phys. Chem. C} \textbf{2014},
  \emph{118}, 11752--11762\relax
\mciteBstWouldAddEndPuncttrue
\mciteSetBstMidEndSepPunct{\mcitedefaultmidpunct}
{\mcitedefaultendpunct}{\mcitedefaultseppunct}\relax
\EndOfBibitem
\bibitem[Cheng and Sprik(2012)Cheng, and Sprik]{Cheng:2012cj}
Cheng,~J.; Sprik,~M. {Alignment of electronic energy levels at electrochemical
  interfaces}. \emph{Phys. Chem. Chem. Phys.} \textbf{2012}, \emph{14},
  11245--11267\relax
\mciteBstWouldAddEndPuncttrue
\mciteSetBstMidEndSepPunct{\mcitedefaultmidpunct}
{\mcitedefaultendpunct}{\mcitedefaultseppunct}\relax
\EndOfBibitem
\bibitem[Cheng \latin{et~al.}(2014)Cheng, Liu, VandeVondele, Sulpizi, and
  Sprik]{Cheng:2014jb}
Cheng,~J.; Liu,~X.; VandeVondele,~J.; Sulpizi,~M.; Sprik,~M. {Redox Potentials
  and Acidity Constants from Density Functional Theory Based Molecular
  Dynamics}. \emph{Acc. Chem. Res.} \textbf{2014}, \emph{47}, 3522--3529\relax
\mciteBstWouldAddEndPuncttrue
\mciteSetBstMidEndSepPunct{\mcitedefaultmidpunct}
{\mcitedefaultendpunct}{\mcitedefaultseppunct}\relax
\EndOfBibitem
\bibitem[Pham \latin{et~al.}(2017)Pham, Ping, and Galli]{Pham:2017bu}
Pham,~T.~A.; Ping,~Y.; Galli,~G. {Modelling heterogeneous interfaces for solar
  water splitting}. \emph{Nat. Mater.} \textbf{2017}, \emph{16}, 401--408\relax
\mciteBstWouldAddEndPuncttrue
\mciteSetBstMidEndSepPunct{\mcitedefaultmidpunct}
{\mcitedefaultendpunct}{\mcitedefaultseppunct}\relax
\EndOfBibitem
\bibitem[Ambrosio \latin{et~al.}(2018)Ambrosio, Wiktor, and
  Pasquarello]{Ambrosio:2018gf}
Ambrosio,~F.; Wiktor,~J.; Pasquarello,~A. {pH-Dependent Catalytic Reaction
  Pathway for Water Splitting at the BiVO 4{\textendash}Water Interface from
  the Band Alignment}. \emph{ACS Energy Lett.} \textbf{2018}, \emph{3},
  829--834\relax
\mciteBstWouldAddEndPuncttrue
\mciteSetBstMidEndSepPunct{\mcitedefaultmidpunct}
{\mcitedefaultendpunct}{\mcitedefaultseppunct}\relax
\EndOfBibitem
\bibitem[Stengel \latin{et~al.}(2009)Stengel, Spaldin, and
  Vanderbilt]{Stengel:2009cd}
Stengel,~M.; Spaldin,~N.~A.; Vanderbilt,~D. {Electric displacement as the
  fundamental variable in electronic-structure calculations}. \emph{Nat. Phys.}
  \textbf{2009}, \emph{5}, 304--308\relax
\mciteBstWouldAddEndPuncttrue
\mciteSetBstMidEndSepPunct{\mcitedefaultmidpunct}
{\mcitedefaultendpunct}{\mcitedefaultseppunct}\relax
\EndOfBibitem
\bibitem[Zhang and Sprik(2016)Zhang, and Sprik]{Zhang:2016cl}
Zhang,~C.; Sprik,~M. {Computing the dielectric constant of liquid water at
  constant dielectric displacement}. \emph{Phys. Rev. B} \textbf{2016},
  \emph{93}, 144201\relax
\mciteBstWouldAddEndPuncttrue
\mciteSetBstMidEndSepPunct{\mcitedefaultmidpunct}
{\mcitedefaultendpunct}{\mcitedefaultseppunct}\relax
\EndOfBibitem
\bibitem[Zhang \latin{et~al.}(2016)Zhang, Hutter, and Sprik]{Zhang:2016ho}
Zhang,~C.; Hutter,~J.; Sprik,~M. {Computing the Kirkwood g-Factor by Combining
  Constant Maxwell Electric Field and Electric Displacement Simulations:
  Application to the Dielectric Constant of Liquid Water}. \emph{J. Phys. Chem.
  Lett.} \textbf{2016}, \emph{7}, 2696--2701\relax
\mciteBstWouldAddEndPuncttrue
\mciteSetBstMidEndSepPunct{\mcitedefaultmidpunct}
{\mcitedefaultendpunct}{\mcitedefaultseppunct}\relax
\EndOfBibitem
\bibitem[Zhang and Sprik(2016)Zhang, and Sprik]{Zhang:2016ca}
Zhang,~C.; Sprik,~M. {Finite field methods for the supercell modeling of
  charged insulator/electrolyte interfaces}. \emph{Phys. Rev. B} \textbf{2016},
  \emph{94}, 245309\relax
\mciteBstWouldAddEndPuncttrue
\mciteSetBstMidEndSepPunct{\mcitedefaultmidpunct}
{\mcitedefaultendpunct}{\mcitedefaultseppunct}\relax
\EndOfBibitem
\bibitem[Sayer \latin{et~al.}(2017)Sayer, Zhang, and Sprik]{Sayer:2017cw}
Sayer,~T.; Zhang,~C.; Sprik,~M. {Charge compensation at the interface between
  the polar NaCl(111) surface and a NaCl aqueous solution}. \emph{J. Chem.
  Phys.} \textbf{2017}, \emph{147}, 104702--8\relax
\mciteBstWouldAddEndPuncttrue
\mciteSetBstMidEndSepPunct{\mcitedefaultmidpunct}
{\mcitedefaultendpunct}{\mcitedefaultseppunct}\relax
\EndOfBibitem
\bibitem[Zhang(2018)]{Zhang:2018cf}
Zhang,~C. {Communication: Computing the Helmholtz capacitance of charged
  insulator-electrolyte interfaces from the supercell polarization}. \emph{J.
  Chem. Phys.} \textbf{2018}, \emph{149}, 031103--6\relax
\mciteBstWouldAddEndPuncttrue
\mciteSetBstMidEndSepPunct{\mcitedefaultmidpunct}
{\mcitedefaultendpunct}{\mcitedefaultseppunct}\relax
\EndOfBibitem
\bibitem[Sayer \latin{et~al.}(2019)Sayer, Sprik, and Zhang]{Sayer:2019bm}
Sayer,~T.; Sprik,~M.; Zhang,~C. {Finite electric displacement simulations of
  polar ionic solid-electrolyte interfaces: Application to NaCl(111)/aqueous
  NaCl solution}. \emph{J. Chem. Phys.} \textbf{2019}, \emph{150},
  041716--13\relax
\mciteBstWouldAddEndPuncttrue
\mciteSetBstMidEndSepPunct{\mcitedefaultmidpunct}
{\mcitedefaultendpunct}{\mcitedefaultseppunct}\relax
\EndOfBibitem
\bibitem[Perdew \latin{et~al.}(1996)Perdew, Burke, and
  Ernzerhof]{PhysRevLett.77.3865}
Perdew,~J.~P.; Burke,~K.; Ernzerhof,~M. {Generalized Gradient Approximation
  Made Simple}. \emph{Phys. Rev. Lett.} \textbf{1996}, \emph{77},
  3865--3868\relax
\mciteBstWouldAddEndPuncttrue
\mciteSetBstMidEndSepPunct{\mcitedefaultmidpunct}
{\mcitedefaultendpunct}{\mcitedefaultseppunct}\relax
\EndOfBibitem
\bibitem[Hutter \latin{et~al.}(2013)Hutter, Iannuzzi, Schiffmann, and
  VandeVondele]{Hutter:2013iea}
Hutter,~J.; Iannuzzi,~M.; Schiffmann,~F.; VandeVondele,~J. {CP2K:atomistic
  simulations of condensed matter systems}. \emph{WIREs Comput. Mol. Sci.}
  \textbf{2013}, \emph{4}, 15--25\relax
\mciteBstWouldAddEndPuncttrue
\mciteSetBstMidEndSepPunct{\mcitedefaultmidpunct}
{\mcitedefaultendpunct}{\mcitedefaultseppunct}\relax
\EndOfBibitem
\bibitem[si()]{si}
See the Supplemental Information for the description of method and
  computational setup.\relax
\mciteBstWouldAddEndPunctfalse
\mciteSetBstMidEndSepPunct{\mcitedefaultmidpunct}
{}{\mcitedefaultseppunct}\relax
\EndOfBibitem
\bibitem[Limmer \latin{et~al.}(2013)Limmer, Merlet, Salanne, Chandler, Madden,
  van Roij, and Rotenberg]{Limmer:2013ir}
Limmer,~D.~T.; Merlet,~C.; Salanne,~M.; Chandler,~D.; Madden,~P.~A.; van
  Roij,~R.; Rotenberg,~B. {Charge Fluctuations in Nanoscale Capacitors}.
  \emph{Phys. Rev. Lett.} \textbf{2013}, \emph{111}, 106102\relax
\mciteBstWouldAddEndPuncttrue
\mciteSetBstMidEndSepPunct{\mcitedefaultmidpunct}
{\mcitedefaultendpunct}{\mcitedefaultseppunct}\relax
\EndOfBibitem
\bibitem[Cheng and Sprik(2014)Cheng, and Sprik]{Cheng:2014eh}
Cheng,~J.; Sprik,~M. {The electric double layer at a rutile TiO2 water
  interface modelled using density functional theory based molecular dynamics
  simulation}. \emph{J. Phys-Condens. Mat.} \textbf{2014}, \emph{26},
  244108\relax
\mciteBstWouldAddEndPuncttrue
\mciteSetBstMidEndSepPunct{\mcitedefaultmidpunct}
{\mcitedefaultendpunct}{\mcitedefaultseppunct}\relax
\EndOfBibitem
\bibitem[Parez \latin{et~al.}(2014)Parez, P{\v{r}}edota, and
  Machesky]{Parez:2014dx}
Parez,~S.; P{\v{r}}edota,~M.; Machesky,~M. {Dielectric Properties of Water at
  Rutile and Graphite Surfaces: Effect of Molecular Structure}. \emph{J. Phys.
  Chem. C} \textbf{2014}, \emph{118}, 4818--4834\relax
\mciteBstWouldAddEndPuncttrue
\mciteSetBstMidEndSepPunct{\mcitedefaultmidpunct}
{\mcitedefaultendpunct}{\mcitedefaultseppunct}\relax
\EndOfBibitem
\bibitem[Creazzo \latin{et~al.}(2019)Creazzo, Galimberti, Pezzotti, and
  Gaigeot]{Creazzo:2019ek}
Creazzo,~F.; Galimberti,~D.~R.; Pezzotti,~S.; Gaigeot,~M.-P. {DFT-MD of the
  (110)-Co3O4 cobalt oxide semiconductor in contact with liquid water,
  preliminary chemical and physical insights into the electrochemical
  environment}. \emph{J. Chem. Phys.} \textbf{2019}, \emph{150},
  041721--19\relax
\mciteBstWouldAddEndPuncttrue
\mciteSetBstMidEndSepPunct{\mcitedefaultmidpunct}
{\mcitedefaultendpunct}{\mcitedefaultseppunct}\relax
\EndOfBibitem
\bibitem[M_z()]{M_z}
Here we plot the supercell dipole moment $M_z$ instead $P_z$ where they differ
  just by the volume $\Omega$ of the model system $M_z=\Omega P_z$.\relax
\mciteBstWouldAddEndPunctfalse
\mciteSetBstMidEndSepPunct{\mcitedefaultmidpunct}
{}{\mcitedefaultseppunct}\relax
\EndOfBibitem
\bibitem[Junquera \latin{et~al.}(2007)Junquera, Cohen, and
  Rabe]{Junquera:2007eb}
Junquera,~J.; Cohen,~M.~H.; Rabe,~K.~M. {Nanoscale smoothing and the analysis
  of interfacial charge and dipolar densities}. \emph{J. Phys-Condens. Mat.}
  \textbf{2007}, \emph{19}, 213203\relax
\mciteBstWouldAddEndPuncttrue
\mciteSetBstMidEndSepPunct{\mcitedefaultmidpunct}
{\mcitedefaultendpunct}{\mcitedefaultseppunct}\relax
\EndOfBibitem
\bibitem[Ong \latin{et~al.}(1992)Ong, Zhao, and Eisenthal]{Ong:1992ca}
Ong,~S.; Zhao,~X.; Eisenthal,~K.~B. {Polarization of water molecules at a
  charged interface: second harmonic studies of the silica/water interface}.
  \emph{Chem. Phys. Lett.} \textbf{1992}, \emph{191}, 327--335\relax
\mciteBstWouldAddEndPuncttrue
\mciteSetBstMidEndSepPunct{\mcitedefaultmidpunct}
{\mcitedefaultendpunct}{\mcitedefaultseppunct}\relax
\EndOfBibitem
\bibitem[Brown \latin{et~al.}(2016)Brown, Abbas, Kleibert, Green, Goel, May,
  and Squires]{Brown:2016bo}
Brown,~M.~A.; Abbas,~Z.; Kleibert,~A.; Green,~R.~G.; Goel,~A.; May,~S.;
  Squires,~T.~M. {Determination of Surface Potential and Electrical
  Double-Layer Structure at the Aqueous Electrolyte-Nanoparticle Interface}.
  \emph{Phys. Rev. X} \textbf{2016}, \emph{6}, 011007--12\relax
\mciteBstWouldAddEndPuncttrue
\mciteSetBstMidEndSepPunct{\mcitedefaultmidpunct}
{\mcitedefaultendpunct}{\mcitedefaultseppunct}\relax
\EndOfBibitem
\bibitem[Bourikas \latin{et~al.}(2001)Bourikas, Hiemstra, and
  Van~Riemsdijk]{Bourikas:2001hh}
Bourikas,~K.; Hiemstra,~T.; Van~Riemsdijk,~W.~H. {Ion Pair Formation and
  Primary Charging Behavior of Titanium Oxide (Anatase and Rutile)}.
  \emph{Langmuir} \textbf{2001}, \emph{17}, 749--756\relax
\mciteBstWouldAddEndPuncttrue
\mciteSetBstMidEndSepPunct{\mcitedefaultmidpunct}
{\mcitedefaultendpunct}{\mcitedefaultseppunct}\relax
\EndOfBibitem
\bibitem[Hiemstra and Van~Riemsdijk(1991)Hiemstra, and
  Van~Riemsdijk]{Hiemstra:1991ju}
Hiemstra,~T.; Van~Riemsdijk,~W.~H. {Physical chemical interpretation of primary
  charging behaviour of metal (hydr) oxides}. \emph{Colloids Surf.}
  \textbf{1991}, \emph{59}, 7--25\relax
\mciteBstWouldAddEndPuncttrue
\mciteSetBstMidEndSepPunct{\mcitedefaultmidpunct}
{\mcitedefaultendpunct}{\mcitedefaultseppunct}\relax
\EndOfBibitem
\bibitem[Yates and Healy(1980)Yates, and Healy]{Yates:1980io}
Yates,~D.~E.; Healy,~T.~W. {Titanium dioxide{\textendash}electrolyte interface.
  Part 2.{\textemdash}Surface charge (titration) studies}. \emph{J. Chem. Soc.,
  Faraday Trans. 1} \textbf{1980}, \emph{76}, 9--10\relax
\mciteBstWouldAddEndPuncttrue
\mciteSetBstMidEndSepPunct{\mcitedefaultmidpunct}
{\mcitedefaultendpunct}{\mcitedefaultseppunct}\relax
\EndOfBibitem
\bibitem[Kallay \latin{et~al.}(1994)Kallay, {\v C}oli{\'c}, Fuerstenau, Jang,
  and Matijevi{\'c}]{Kallay:1994iu}
Kallay,~N.; {\v C}oli{\'c},~M.; Fuerstenau,~D.~W.; Jang,~H.~M.;
  Matijevi{\'c},~E. {Lyotropic effect in surface charge, electrokinetics, and
  coagulation of a rutile dispersion}. \emph{Colloid Polym. Sci.}
  \textbf{1994}, \emph{272}, 554--561\relax
\mciteBstWouldAddEndPuncttrue
\mciteSetBstMidEndSepPunct{\mcitedefaultmidpunct}
{\mcitedefaultendpunct}{\mcitedefaultseppunct}\relax
\EndOfBibitem
\bibitem[Blok and de~Bruyn(1970)Blok, and de~Bruyn]{Blok:1970jn}
Blok,~L.; de~Bruyn,~P.~L. {The ionic double layer at the ZnO solution
  interface. III. Comparison of calculated and experimental differential
  capacities}. \emph{J. Colloid Interface Sci.} \textbf{1970}, \emph{32},
  533--538\relax
\mciteBstWouldAddEndPuncttrue
\mciteSetBstMidEndSepPunct{\mcitedefaultmidpunct}
{\mcitedefaultendpunct}{\mcitedefaultseppunct}\relax
\EndOfBibitem
\bibitem[Sverjensky(2005)]{Sverjensky:2005fe}
Sverjensky,~D.~A. {Prediction of surface charge on oxides in salt solutions:
  Revisions for 1:1 (M+L-) electrolytes}. \emph{Geochim. Cosmochim. Acta}
  \textbf{2005}, \emph{69}, 225--257\relax
\mciteBstWouldAddEndPuncttrue
\mciteSetBstMidEndSepPunct{\mcitedefaultmidpunct}
{\mcitedefaultendpunct}{\mcitedefaultseppunct}\relax
\EndOfBibitem
\bibitem[Bankura \latin{et~al.}(2013)Bankura, Carnevale, and
  Klein]{Bankura:2013bd}
Bankura,~A.; Carnevale,~V.; Klein,~M.~L. {Hydration structure of salt solutions
  from ab initio molecular dynamics}. \emph{J. Chem. Phys.} \textbf{2013},
  \emph{138}, 014501\relax
\mciteBstWouldAddEndPuncttrue
\mciteSetBstMidEndSepPunct{\mcitedefaultmidpunct}
{\mcitedefaultendpunct}{\mcitedefaultseppunct}\relax
\EndOfBibitem
\bibitem[Serrano \latin{et~al.}(2015)Serrano, Bonanni, Di~Giovannantonio,
  Kosmala, Schmid, Diebold, Di~Carlo, Cheng, VandeVondele, Wandelt, and
  Goletti]{Serrano:2015it}
Serrano,~G.; Bonanni,~B.; Di~Giovannantonio,~M.; Kosmala,~T.; Schmid,~M.;
  Diebold,~U.; Di~Carlo,~A.; Cheng,~J.; VandeVondele,~J.; Wandelt,~K.
  \latin{et~al.}  {Molecular Ordering at the Interface Between Liquid Water and
  Rutile TiO2(110)}. \emph{Adv. Mater. Interfaces} \textbf{2015}, \emph{2},
  1500246--6\relax
\mciteBstWouldAddEndPuncttrue
\mciteSetBstMidEndSepPunct{\mcitedefaultmidpunct}
{\mcitedefaultendpunct}{\mcitedefaultseppunct}\relax
\EndOfBibitem
\bibitem[Kirkwood(1939)]{Kirkwood:1939br}
Kirkwood,~J.~G. {The Dielectric Polarization of Polar Liquids}. \emph{J. Chem.
  Phys.} \textbf{1939}, \emph{7}, 911--919\relax
\mciteBstWouldAddEndPuncttrue
\mciteSetBstMidEndSepPunct{\mcitedefaultmidpunct}
{\mcitedefaultendpunct}{\mcitedefaultseppunct}\relax
\EndOfBibitem
\bibitem[Imanishi and Fukui(2014)Imanishi, and Fukui]{Imanishi:2014cj}
Imanishi,~A.; Fukui,~K.-i. {Atomic-Scale Surface Local Structure of TiO2 and
  Its Influence on the Water Photooxidation Process}. \emph{J. Phys. Chem.
  Lett} \textbf{2014}, \emph{5}, 2108--2117\relax
\mciteBstWouldAddEndPuncttrue
\mciteSetBstMidEndSepPunct{\mcitedefaultmidpunct}
{\mcitedefaultendpunct}{\mcitedefaultseppunct}\relax
\EndOfBibitem
\end{mcitethebibliography}

\providecommand{\latin}[1]{#1}
\makeatletter
\providecommand{\doi}
  {\begingroup\let\do\@makeother\dospecials
  \catcode`\{=1 \catcode`\}=2 \doi@aux}
\providecommand{\doi@aux}[1]{\endgroup\texttt{#1}}
\makeatother
\providecommand*\mcitethebibliography{\thebibliography}
\csname @ifundefined\endcsname{endmcitethebibliography}
  {\let\endmcitethebibliography\endthebibliography}{}

\end{document}


\maketitle

\makeatletter 
\renewcommand{\thefigure}{S\@arabic\c@figure} 
\renewcommand{\thetable}{S\@arabic\c@table} 
\makeatother

\appendix{} 
\setcounter{figure}{0} 
\setcounter{table}{0}

\centerline{\bf\Large{Supporting Information}}

\section{\label{sec:level1} The hybrid constant $\bar{D}$ Hamiltonian}

In this study, we used the hybrid SSV constant $\bar{D}$ Hamiltonian which can be derived either from the thermodynamics
argument originally~\cite{Stengel:2009cd} or from a current dependent
Lagrangian as shown recently~\cite{Sprik:2018kr}:
\begin{equation}
\label{uvdb}
H_D\left(v , \varphi, \bar{D} \right) = H_{\mathrm{PBC}}(v, \varphi)+ \frac{\mathit{\Omega}}{8 \pi}
  \left(\bar{D} - 4 \pi P_z(v,\varphi) \right)^2
\end{equation}
where $P_z$ is the itinerant polarization in the direction of
$\bar{D}$ (See
Secs. IV B and IV C in Ref. ~\cite{Zhang:2016ca} for the elaboration), which is formally
defined as a time integral of the volume integral of current ~\cite{King-Smith:1993prb,Resta:1994rmp,Resta:2007ch,
  Caillol:1989jcpa,Caillol:1989jcpb,Caillol:1994ho}. $\Omega$ is the supercell
volume and $v= (\mathbf{r}^N ,\mathbf{p}^N)$ are the classical degrees
of freedom of a $N$-particle system. The electronic degrees of freedom
$\varphi_\alpha, \; \alpha =1 \dots M$  determining the induced
dipoles are collectively represented by $\varphi$.  The bar over $D$
emphasizes that it is a controll variable instead of
an observable. ``Hybrid'' means the field is only applied in the
direction perpendicular to the surface.

In CP2K, one can switch on the hybrid constant $\bar{D}=0$ scheme by
using the following keyword block:

\begin{lstlisting} 
&PERIODIC_EFIELD
     INTENSITY 0.00
     DISPLACEMENT_FIELD T
     POLARISATION 0.0 0.0 1.0
     D_FILTER 0.0 0.0 1.0
&END PERIODIC_EFIELD
 \end{lstlisting}
 
\section{\label{sec:level1} CP2K simulations of electrified rutile TiO2(110)-NaCl
  electrolyte interfaces}

The electronic structure of rutile TiO$_2$ (110)-NaCl electrolyte systems was solved applying DFT in the  Perdew-Burke-Ernzerhof (PBE) approximation~\cite{PhysRevLett.77.3865} using CP2K suit of programs~\cite{VandeVondele:2005ge, Hutter:2013iea}. 
Basis sets optimized for molecular systems (MOLOPT)~\cite{VandeVondele:2007wt} with double-$\zeta$ basis sets with one additional polarization functions (DZVP) and a charge density cutoff of 320 Ry were used. Core electrons were taken into account using the dual-space Goedecker-Teter-Hutter (GTH) pseudopotentials~\cite{Goedecker:1996ve}. 

The model system consisted of a symmetric periodic slab of five layer
O-Ti-O trilayer with (4x2) surface cell, 110 water molecules, 5 Na$^+$
and 5 Cl$^-$ ions in a periodic cell of
11.90~\AA$\times$13.20~\AA$\times$38.28~\AA. To keep the composition
fixed in all setups, the negatively charged side of TiO$_2$ slab is formed by removing
H$^+$ from absorbed water molecules and the positively charged 
side is formed by adding removed H$^+$ to oxygen sites of the other
side of TiO$_2$ slab. The integration time-step is 0.5 fs and MD trajectories were collected for about 15ps for each step-up at $\bar{D}=0$ condition after initial equilibration at $\bar{E}=0$ condition. The initial configuration for different charge densities were obtained from our previous work of charged insulator-electrolyte systems using a classical point-charge like model~\cite{Zhang:2016ca}. Bussi-Donadio-Parrinello thermostat ~\cite{Bussi:2008wu} was used to keep the temperature at 330K throughout all simulations.  

The macroscopic polarization is computed using both Resta
formula~\cite{RESTA:1998hj} and maximally-localized Wannier functions
(MLWFs)~\cite{Marzari:2012eu}, as implemented in CP2K. This is an
important technical check to ensure the branch consistency because of the
multi-valued nature of the macroscopic polarization, as discussed in
details in Sec. VI B in Ref.~\cite{Sayer:2019bm}. Electrostatic potential cube files were saved from
CP2K every 50 fs and macro-averaging of the averaged
electrostatic potential cube file was carried out using MACROAVE code~\cite{Junquera:2007eb}. 
